\documentclass[]{spie}  

\usepackage{xcolor}
\usepackage{amsmath,amsfonts,amssymb}
\usepackage{graphicx}
\usepackage[colorlinks=true, allcolors=blue]{hyperref}
\usepackage{siunitx}
\usepackage{makecell}
\usepackage[acronym]{glossaries}
\usepackage[nolist]{acronym}
\newacro{ADC}[ADC]{Atmospheric Dispersion Corrector}
\newacro{SAF}[SAF]{Science Archive Facility}
\newacro{SOXS}[SOXS]{Son-Of-X-shooter}
\newacro{NTT}[NTT]{New Technology Telescope}
\newacro{ESO}[ESO]{European Southern Observatory}
\newacro{DRAWER}[DRAWER]{Data Reduction And WEllness Reporter}
\newacro{AC}[AC]{Acquisition and Imaging Camera}
\newacro{CPL}[CPL]{Common Pipeline Library}
\newacro{CLI}[CLI]{Command-Line Interface}
\newacro{QC}[QC]{Quality Control}
\newacro{E2E}[E2E]{End-to-End}

\usepackage{color}
\usepackage{colortbl}
\usepackage{multirow}
\usepackage{listings}
\usepackage{float}
\usepackage{color}
\usepackage{xcolor,colortbl}
\usepackage{caption}
\usepackage{subcaption}
\definecolor{red}{HTML}{dc322f}
\definecolor{green}{HTML}{859900}
\definecolor{blue}{HTML}{268bd2}
\definecolor{orange}{HTML}{cb4b16}
\definecolor{cyan}{HTML}{2aa198}
\definecolor{magneta}{HTML}{d33682}
\definecolor{violet}{HTML}{6c71c4}

\definecolor{dkgreen}{rgb}{0,0.6,0}
\definecolor{gray}{rgb}{0.5,0.5,0.5}
\definecolor{mauve}{rgb}{0.58,0,0.82}

\title{The \ac{SOXS} Data-Reduction Pipeline}

\author[a]{David R. Young}
\author[b]{Marco Landoni}
\author[a]{Stephen J. Smartt}
\author[b]{Sergio~Campana}
\author[b]{Paolo~D'Avanzo}
\author[s]{Riccardo~Claudi}
\author[d]{Pietro~Schipani}
\author[b]{Matteo~Aliverti}
\author[s]{Andrea~Baruffolo}
\author[h,e]{Sagi~Ben-Ami}
\author[d]{Giulio~Capasso}
\author[f,k]{Rosario~Cosentino}
\author[g]{Francesco~D'Alessio}
\author[h]{Ofir	Hershko}
\author[j,q]{Hanindyo~Kuncarayakti}
\author[k]{Matteo~Munari}
\author[m,t]{Giuliano~Pignata}
\author[s]{Kalyan~Radhakrishnan}
\author[c]{Adam~Rubin}
\author[v,k]{Salvatore~Scuderi}
\author[g]{Fabrizio~Vitali}
\author[l]{Jani~Achrén}
\author[w]{José~Antonio~Araiza-Duran}
\author[n]{Iair~Arcavi}
\author[s]{Federico~Battaini}
\author[w]{Anna~Brucalassi}
\author[h]{Rachel~Bruch}
\author[s]{Enrico~Cappellaro}
\author[d]{Mirko~Colapietro}
\author[d]{Massimo~Della~Valle}
\author[k]{Rosario~Di~Benedetto}
\author[d]{Sergio~D'Orsi}
\author[h]{Avishay~Gal-Yam}
\author[b]{Matteo~Genoni}
\author[f]{Marcos~Hernandez}
\author[j,q]{Jari~Kotilainen}
\author[r]{Gianluca~Li~Causi}
\author[d]{Laurent~Marty}
\author[q]{Seppo~Mattila}
\author[h]{Michael~Rappaport}
\author[s]{Davide~Ricci}
\author[b]{Marco~Riva}
\author[s]{Bernardo~Salasnich}
\author[k]{Ricardo~Zanmar~Sanchez}
\author[u]{Maximilian~Stritzinger}
\author[f]{Hector~Ventura}

\affil[a]{Astrophysics Research Centre, School of Mathematics and Physics, Queen's University Belfast, Belfast BT7 1NN, UK}
\affil[b]{INAF -- Osservatorio Astronomico di Brera, Via Bianchi 46, I-23807, Merate, Italy }
\affil[c]{ESO, Karl Schwarzschild Strasse 2, D-85748, Garching bei München, Germany }
\affil[d]{INAF -- Osservatorio Astronomico di Capodimonte, Sal. Moiariello 16, I-80131, Naples, Italy }
\affil[e]{Harvard-Smithsonian Center for Astrophysics, Cambridge, USA }
\affil[f]{FGG-INAF, TNG, Rambla J.A. Fernández Pérez 7, E-38712 Breña Baja (TF), Spain }
\affil[g]{INAF -- Osservatorio Astronomico di Roma, Via Frascati 33, I-00078 M. Porzio Catone, Italy }
\affil[h]{Weizmann Institute of Science, Herzl St 234, Rehovot, 7610001, Israel }
\affil[i]{Max-Planck-Institut für Extraterrestrische Physik, Giessenbachstr. 1, D-85748 Garching, Germany }
\affil[j]{Finnish Centre for Astronomy with ESO (FINCA), FI-20014 University of Turku, Finland}
\affil[k]{INAF -- Osservatorio Astrofisico di Catania, Via S. Sofia 78 30, I-95123 Catania, Italy }
\affil[l]{Incident Angle Oy, Capsiankatu 4 A 29, FI-20320 Turku, Finland }
\affil[m]{Universidad Andres Bello, Avda. Republica 252, Santiago, Chile }
\affil[n]{Tel Aviv University, Department of Astrophysics, 69978 Tel Aviv, Israel }
\affil[o]{Dark Cosmology Centre, Juliane Maries Vej 30, DK-2100 Copenhagen, Denmark }
\affil[p]{Aboa Space Research Oy, Tierankatu 4B, FI-20520 Turku, Finland}
\affil[q]{Tuorla Observatory, Dept. of Physics and Astronomy, FI-20014 University of Turku, Finland }
\affil[r]{INAF - Istituto di Astrofisica e Planetologia Spaziali, Rome, Italy}
\affil[s]{INAF -- Osservatorio Astronomico di Padova, Vicolo dell’Osservatorio 5, I-35122, Padua, Italy }
\affil[t]{Millennium Institute of Astrophysics (MAS)}
\affil[u]{Aarhus University, Ny Munkegade 120, D-8000 Aarhus, Denmark }
\affil[v]{INAF - Istituto di Astrofisica Spaziale e Fisica Cosmica, Via Corti 12, I-20133, Milano, Italy }
\affil[w]{INAF-Osservatorio Astrofisico di Arcetri, Largo E. Fermi 5, I-50125, Firenze, Italy }

\authorinfo{Further author information: (Send correspondence to David Young)\\David Young: E-mail: d.r.young@qub.ac.uk\\  Marco Landoni: E-mail: marco.landoni@inaf.it\\Stephen Smartt: E-mail:  s.smartt@qub.ac.uk}

\begin{document} 
\maketitle

\begin{abstract}
The Son-Of-XShooter (SOXS) is a single object spectrograph (UV-VIS \& NIR) and acquisition camera scheduled to be mounted on the \ac{ESO} 3.58-m New Technology Telescope at the La Silla Observatory. Although the underlying data reduction processes to convert raw detector data to fully-reduced science ready data are complex and multi-stepped, we have designed the \ac{SOXS} Data Reduction pipeline with the core aims of providing end-users with a simple-to-use, well-documented command-line interface while also allowing the pipeline to be run in a fully automated state; streaming reduced data into the \ac{ESO} \ac{SAF} without need for human intervention. To keep up with the stream of data coming from the instrument, there is the requirement to optimise the software to reduce each observation block of data well within the typical observation exposure time. The pipeline is written in Python 3 and has been built with an agile development philosophy that includes CI and adaptive planning.
\end{abstract}

\keywords{SOXS, Pipeline, Data Reduction, Spectroscopy, Imaging}

\section{INTRODUCTION}
\label{sec:intro}  

The \ac{SOXS} (Son Of X-Shooter) instrument is a new medium resolution spectrograph ($R\simeq4500$) capable of simultaneously observing 350-2000nm (U- to H-band) to a limiting magnitude of R $\sim 20$ (3600sec, S/N $\sim$ 10). It shall be hosted at the Nasmyth focus of the \ac{NTT} at La Silla Observatory, Chile (see \cite{Schipani20} for an overview). This paper describes the design of the \ac{SOXS} data-reduction pipeline and data-flow system. Details of each of the other \ac{SOXS} subsystems can be found in a set of related \cite{Aliverti18,Aliverti20,Biondi18,Biondi20,Brucalassi18,Brucalassi20,Capasso18,Claudi18,Claudi20,Colapietro20,Cosentino18,Cosentino20,Genoni20,Kuncarayakti20,Ricci18,Ricci20,Rubin18,Rubin20,Sanchez18,Sanchez20,Schipani16,Schipani18,Schipani20,Vitali18,Vitali20,Young20,Marty22,Landoni22}.

Details of the three detectors included in the \ac{SOXS} instrument, that the pipeline shall receive data from, are given in Section \ref{sec:detectors}. Section \ref{sec:remit} explains the aims and goals of the pipeline and Section \ref{sec:architecture} describes the pipeline software architecture and development environment. Finally, Section \ref{sec:dataflow} outlines the data-products to be expected from the pipeline and the planned data-flow from raw data coming off the telescope through to the data owners collecting reduced data from the \ac{ESO} \ac{SAF}.

\section{The 3 \ac{SOXS} Detectors}
\label{sec:detectors} 

\ac{SOXS} comprises three instruments; the UV-VIS and NIR spectrographs and an \ac{AC}. The instruments are to be mounted on the NTT's Nasmyth focus rotator flange. It is the role of the \ac{SOXS} data reduction pipeline to reduce the pixel data collected by each of these instruments into science-ready data products.

\subsection{The NIR Spectrograph}

\begin{figure}[ht]
  \begin{center}
  \begin{tabular}{c} 
  \includegraphics[width=10cm]{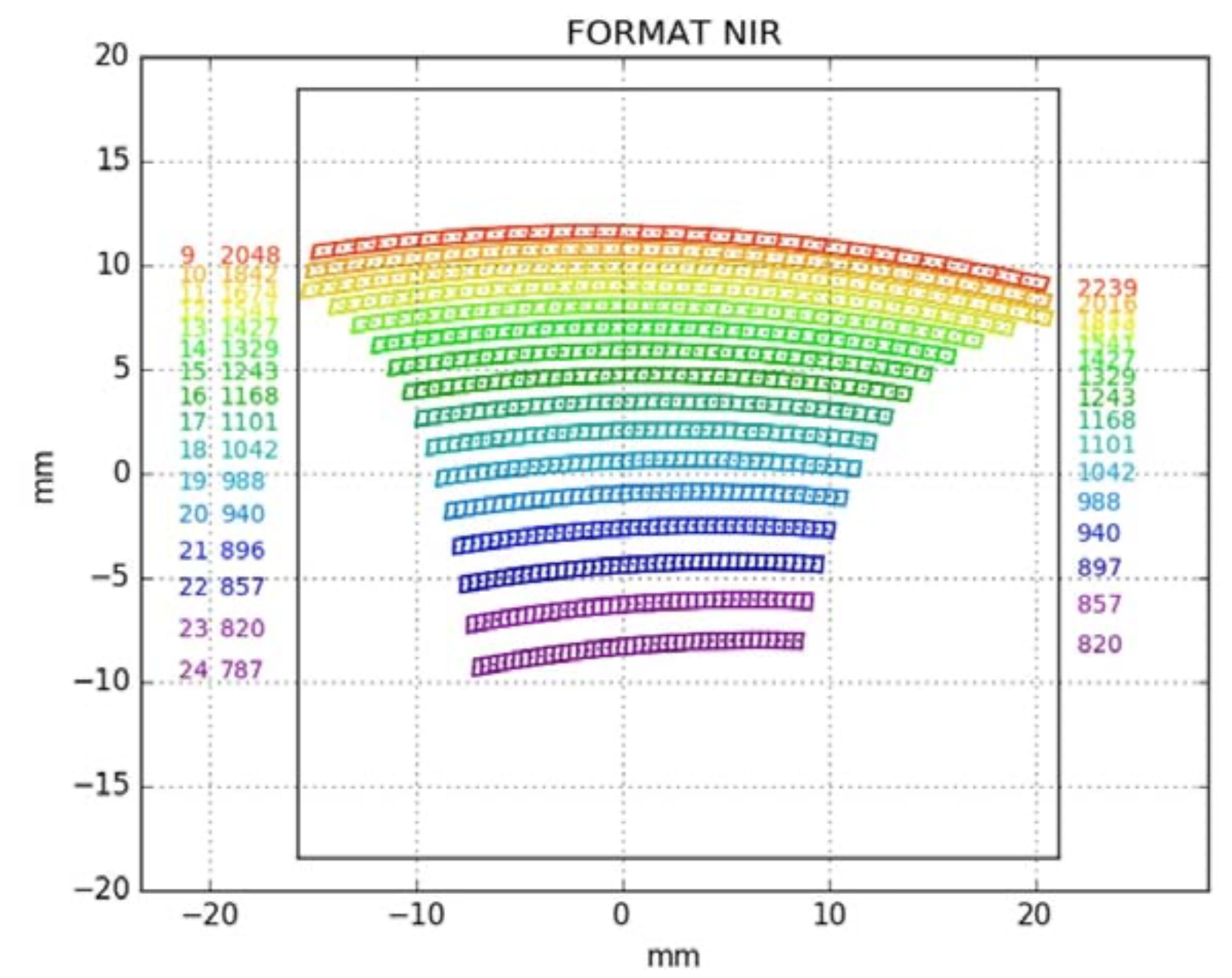}
  \end{tabular}
  \end{center}
  \caption[nir-spectral-format] 
  { \label{fig:nir-spectral-format} 
The \ac{SOXS} NIR spectral format, a reproduction of Figure 5 of Vitali et al. (2018)\cite{Vitali18}. }
  \end{figure}

The \ac{SOXS} NIR spectrograph is a cross-dispersed echelle, employing ``4C' (Collimator Correction of Camera Chromatism) to image spectra in 800-2000nm wavelength range, in 15 orders, onto a 2kx2k 18 micron pixel Teledyne H2RG TM array (see Figure \ref{fig:nir-spectral-format}). It will achieve a spectral resolution R $\simeq$ 5000 (1 arcsec slit).  

\subsection{The UV-VIS Spectrograph}

\begin{figure}[ht]
  \begin{center}
  \begin{tabular}{c} 
  \includegraphics[width=10cm]{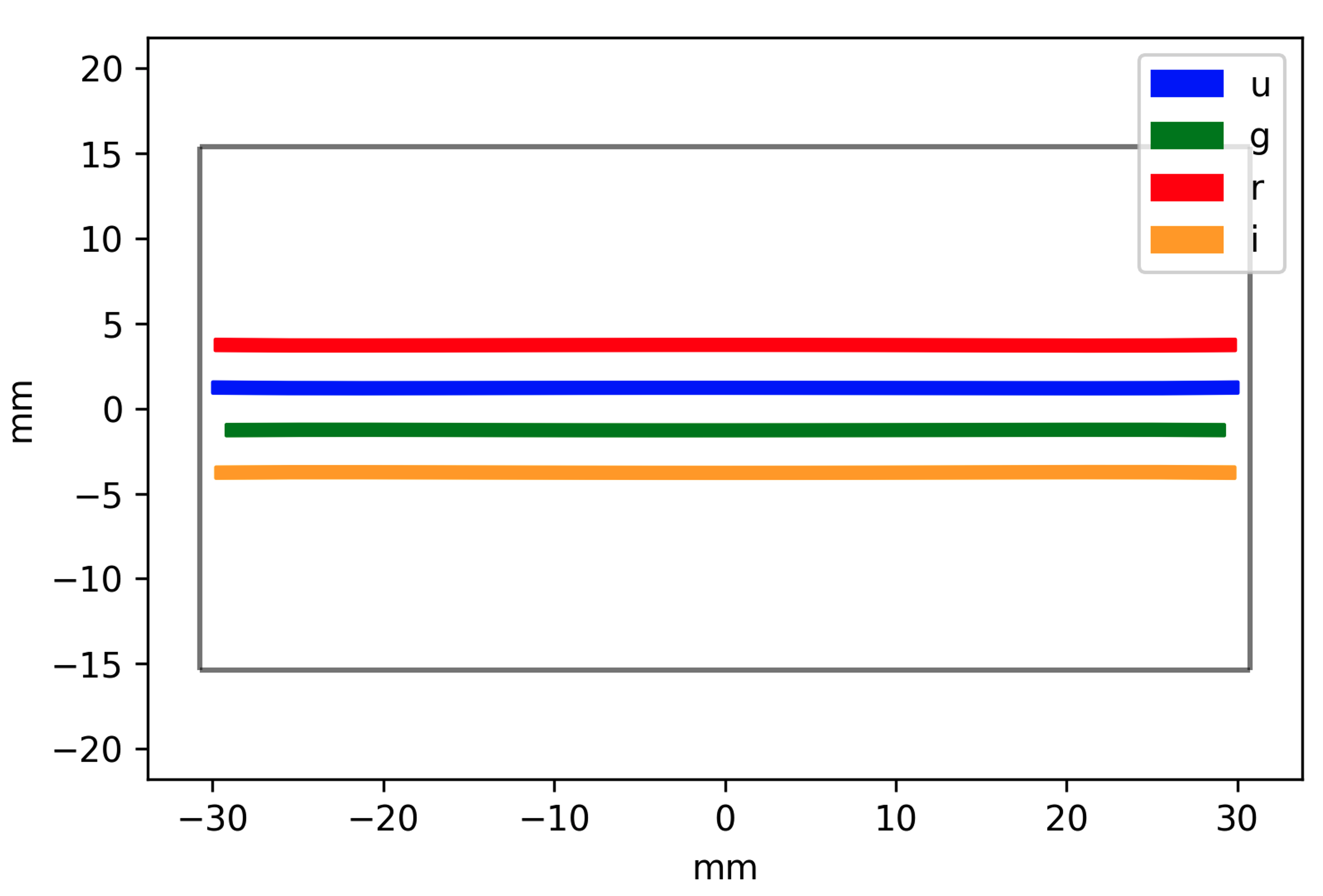}
  \end{tabular}
  \end{center}
  \caption[uvvis-spectral-format] 
  { \label{fig:uvvis-spectral-format} 
The four quasi-orders of the \ac{SOXS} UV-VIS spectral format aligned linearly along the direction of the detector columns.}
  \end{figure}

The UV-VIS spectrograph employs the novel design of 4 ion-etched transmission gratings in the first order ($m=1$) to obtain spectra in the 350-850nm wavelength range (providing an overlap of 50nm with the NIR arm for cross-calibration). The spectral band is split into four poly-chromatic channels and sent to their specific grating \cite{Rubin20}.  Unlike the NIR arm, the UV-VIS arm will include an \ac{ADC}. Each of the four dispersion orders are imaged to separate areas of the e2V CCD and aligned linearly along the direction of the CCD columns (see Figure \ref{fig:uvvis-spectral-format}). 

\subsection{The Acquisition and Imaging Camera}

Although the primary use of the \ac{SOXS} acquisition camera is to acquire spectral targets to allow for their centring on the slit, the camera's $3.5' \times 3.5'$ FOV and 0.205 arcsec/px scale will also allow for science-grade, multi-band imaging. Observers will be able to select from 7 filters; the LSST $u, g, r, i, z, y$ set and Johnson $V$.

\section{Data Reduction Pipeline Remit}
\label{sec:remit} 

The main purpose of the \ac{SOXS} Data Reduction pipeline is to use \ac{SOXS} calibration data (typically, but not necessarily, collected close in time to the science data), to remove all instrument signatures from the \ac{SOXS} scientific data frames, convert this data into physical units and deliver them with their associated error bars to the \ac{ESO} \ac{SAF} as Phase 3 compliant science data products, all within a timescale shorter than a typical \ac{SOXS} science exposure. The pipeline must also support the reduction of data taken in each of the available \ac{SOXS} observation modes. 
The primary reduced pipeline product will be a detrended, wavelength and flux calibrated, telluric corrected 1D spectrum with UV-VIS + NIR arms stitched together (see Section \ref{sec:dataflow}).

\begin{figure}[ht]\centering
  \includegraphics[width=15cm]{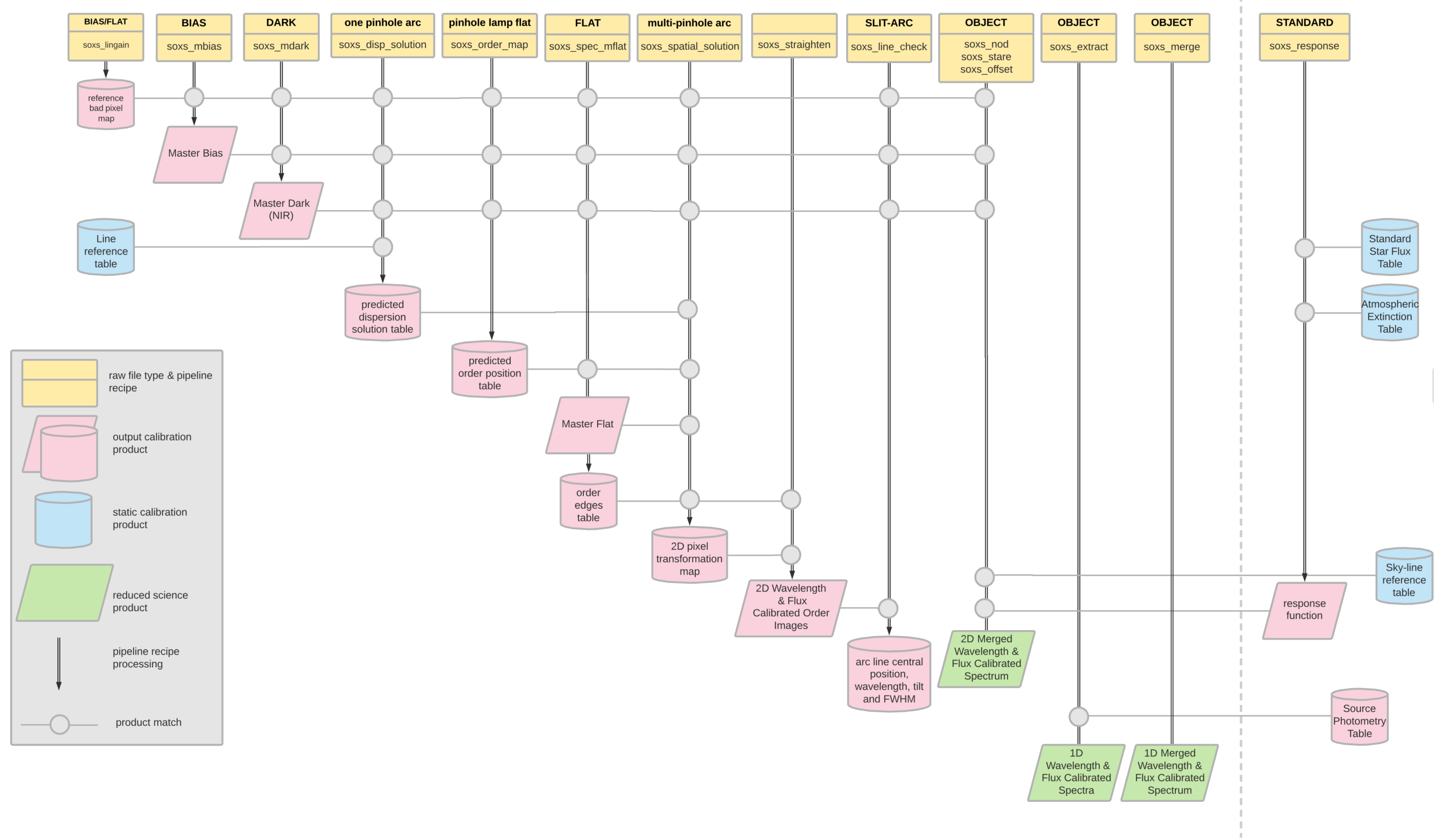}
  \caption{The \ac{SOXS} Spectroscopic Data Reduction Cascade. Each of the vertical lines in the map depicts a raw data frame, the specific recipe to be applied to that frame and the data product(s) output by that recipe. Horizontal lines show how those output data products are used by subsequent pipeline recipes. Time loosely proceeds from left to right (recipe order) and from top-to-bottom (recipe processing steps) on the map.}
  \label{fig:soxs_spectroscopic_data_reduction_cascade}
\end{figure}

\begin{figure}[ht]\centering
  \includegraphics[width=10cm]{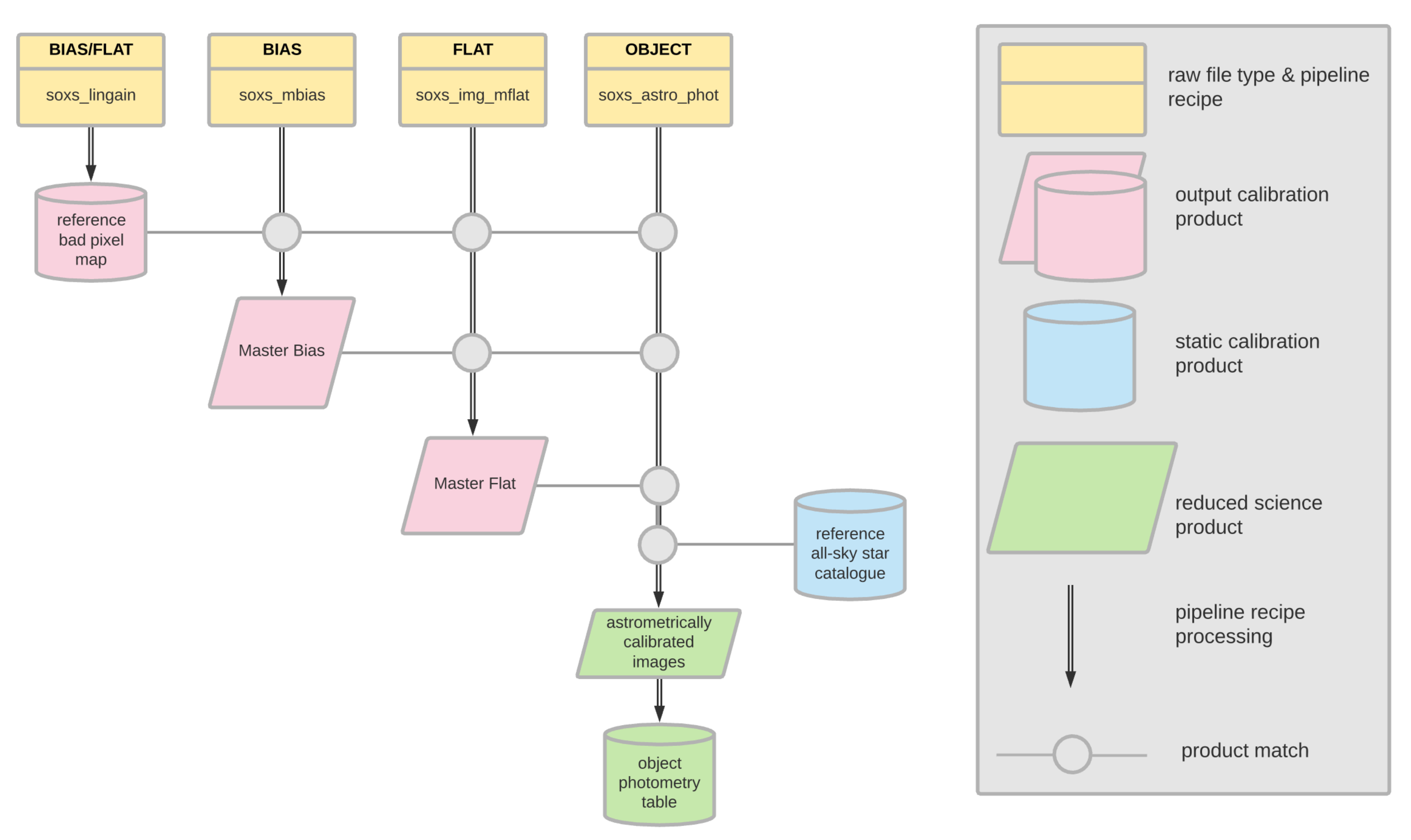}
  \caption{\ac{SOXS} Imaging Data Reduction Cascade. See caption of Figure \ref{fig:soxs_spectroscopic_data_reduction_cascade} for details.}
  \label{fig:soxs_imaging_data_reduction_cascade}
\end{figure}

Although the underlying data reduction processes to convert the raw detector data to fully-reduced, flux- and wavelength-calibrated science-ready data are complex and multi-stepped, \texttt{soxspipe} has been designed with a core aim of providing end-users with an easy-to-install, simple-to-use, clear, well-documented command-line interface while also allowing the pipeline to be run in a fully automated state; streaming reduced \ac{SOXS} data into the \ac{ESO} \ac{SAF} without need for human intervention. Once users have miniconda\footnote{miniconda
  https://docs.conda.io/en/latest/miniconda.html}  or anaconda\footnote{anaconda
  https://www.anaconda.com} installed on their local machine, the pipeline can be installed via a single command and typically takes $<$ 1 min to install.

\begin{lstlisting}
conda create -n soxspipe python=3.8 soxspipe -c conda-forge
\end{lstlisting}

The static calibration files required by the pipeline are shipped alongside the code, removing the burden often required of pipeline users to separately download and manage these files. This has the added benefit of these files being version controlled alongside the code so the end-user will always have access to the suite of calibration files associated with the specific version of the pipeline they have installed on their machine.

The pipeline will also generate \ac{QC} metrics to monitor telescope, instrument and detector health. These metrics are to be read and presented by the \ac{SOXS} health-monitoring system \cite{Marty22}.

\begin{figure}[t]\centering
  \includegraphics[width=10cm]{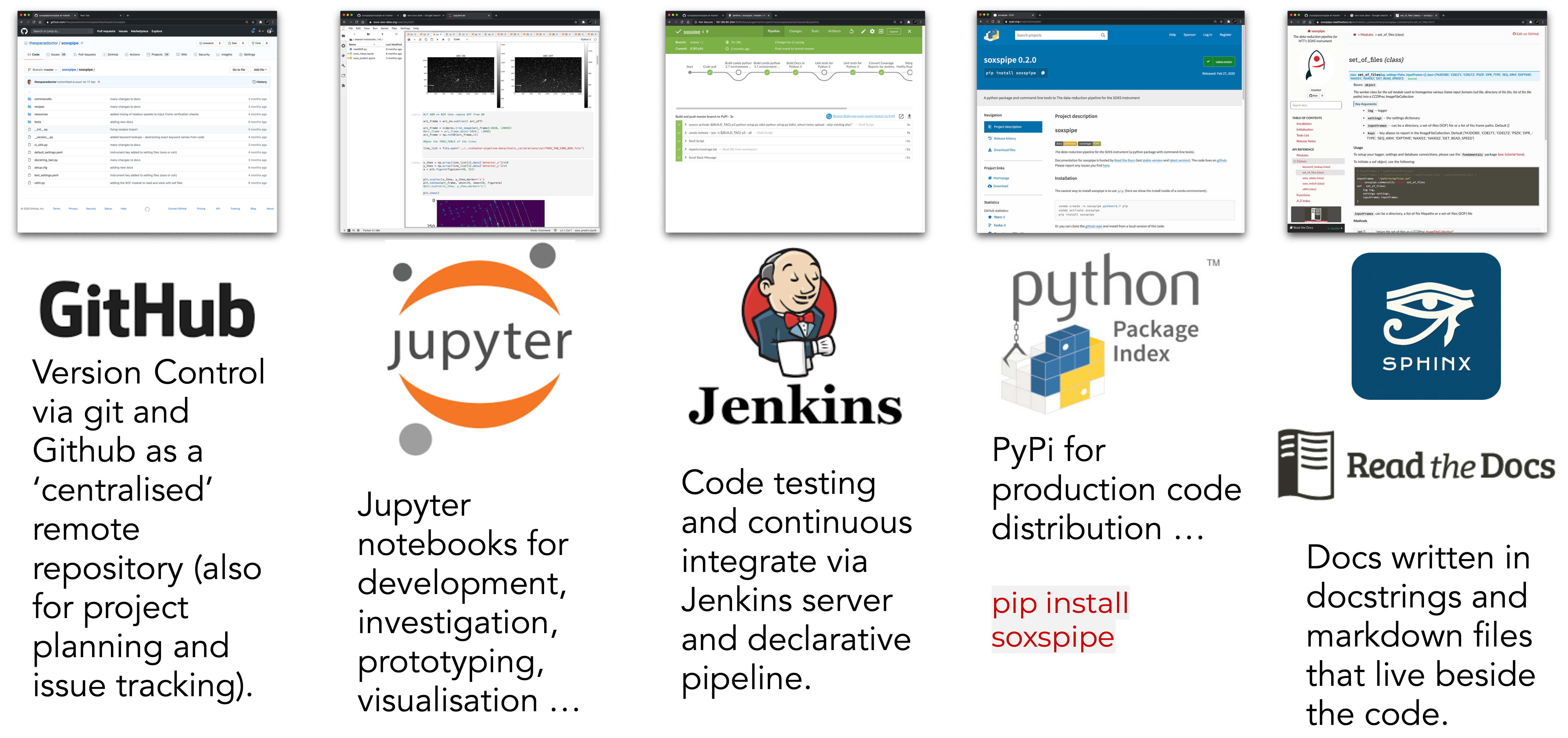}
  \caption{The pipeline development environment. The code is hosted and version controlled on Github, collaborative programming and prototyping is performed within Jupyter notebooks, Jenkins is used for Continuous Integration, code is distributed via PyPi and Conda and documentation is written host on readthedocs.}
  \label{fig:devenv}
\end{figure}

\section{Pipeline Architecture and Development Environment}
\label{sec:architecture}

Presently, the astronomical community have overwhelmingly adopted Python as their scripting language of choice and there are a plethora of well-maintained, mature python packages to help with basic data-reduction routines, visualisation, user-interaction and data manipulation. It was a natural choice therefore to develop the \ac{SOXS} pipeline in Python 3. We have implemented an object-orientated composition and the pipeline is designed to be primarily driven from the command line. The concept of `recipes', originally employed by \ac{ESO}'s \ac{CPL}, has been adopted to define the modular components of the data reduction workflow. These recipes can be connected together to create an end-to-end data-reduction cascade, taking as input raw and calibration frames from the \ac{SOXS} instrument and processing them all the way through to fully reduced, calibrated, \ac{ESO} Phase 3 compliant science products (see Figures \ref{fig:soxs_spectroscopic_data_reduction_cascade} and \ref{fig:soxs_imaging_data_reduction_cascade}). Recipes are named with the prefix `soxs' followed by a succinct description of the recipe (e.g. \texttt{soxs\_mbais} for the master bias creation recipe). There are also many reusable functions designed to be called from multiple recipes; these are referred to as `utilities' in \texttt{soxspipe}. 

The pipeline has been built with an agile development philosophy that includes adaptive planning and evolutionary development. As with any software project, one of the greatest risks is knowledge loss due to a team member leaving before project completion. To mitigate this risk we have employed pair-programming techniques to share knowledge, both explicit and tacit, between two developers. In times of travel bans and remote working a JupterHub server with Python-based notebooks, shared screens and video conferencing tools have been essential to executing these techniques.

\begin{figure}[t]
     \centering
     \begin{subfigure}[b]{0.45\textwidth}
         \centering
         \includegraphics[width=\textwidth]{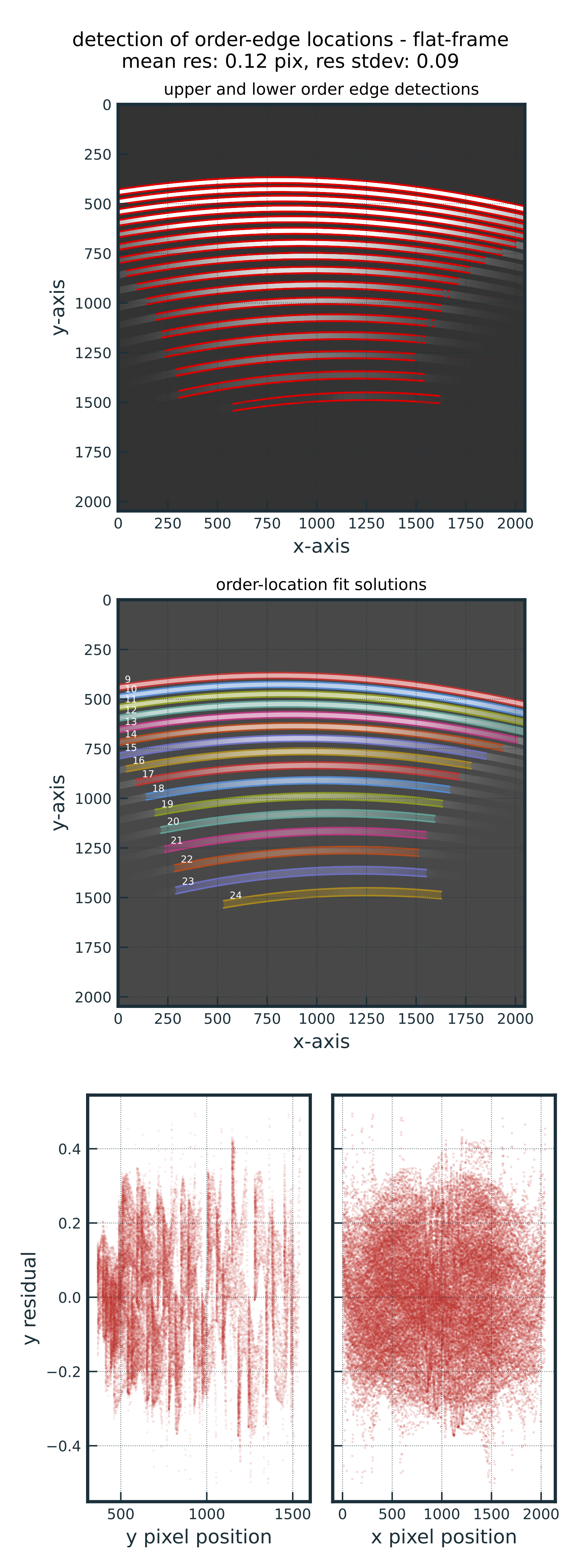}
         \label{fig:order-edge}
     \end{subfigure}
     \hfill
     \begin{subfigure}[b]{0.54\textwidth}
         \centering
         \includegraphics[width=\textwidth]{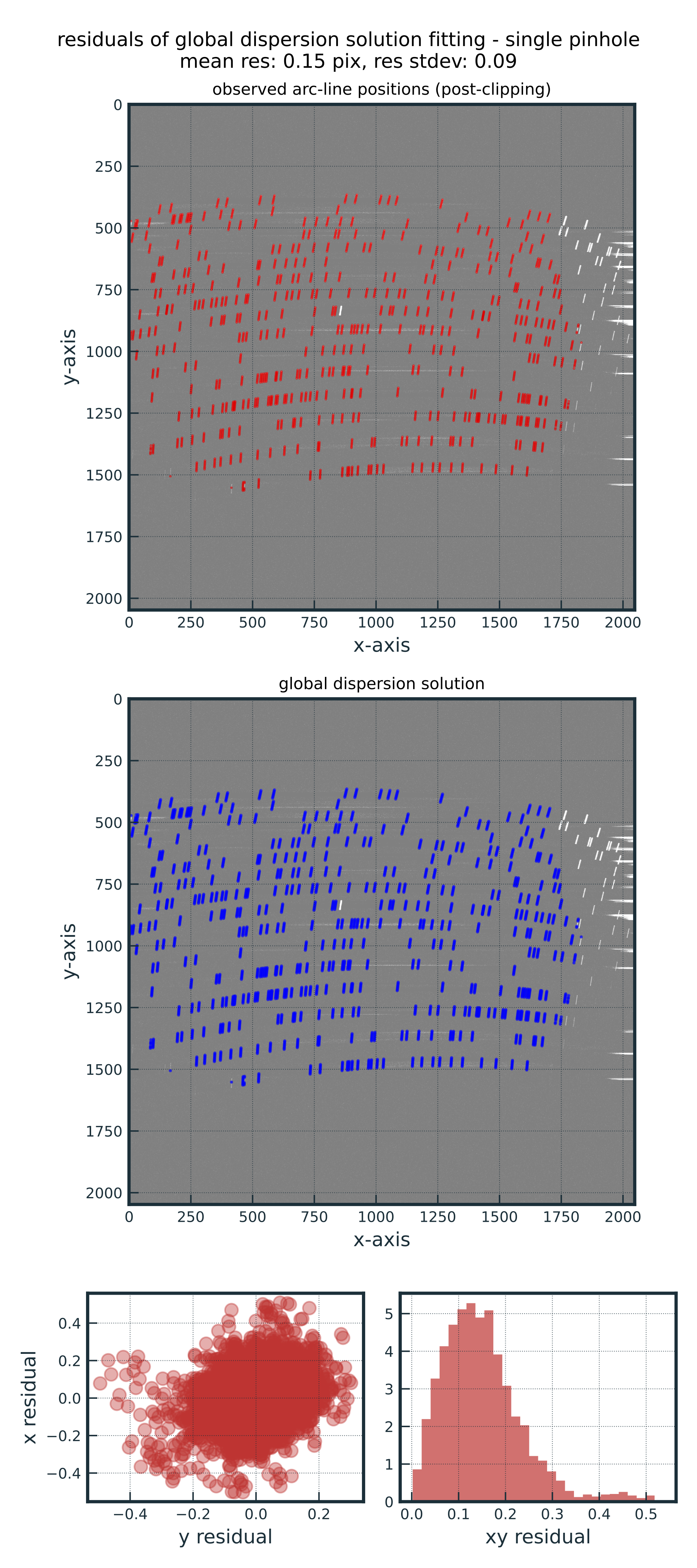}
         \label{fig:dispsol}
     \end{subfigure}
        \caption{The left panels show the NIR order-edges as identified by the \ac{SOXS} data-reduction pipeline using a master-flat frame created from a set of full slit flat-lamp frames generated by the \ac{E2E} simulator. On the right, the resulting final dispersion solution and residuals as fitted by the \ac{SOXS} data-reduction pipeline using a simulated arc-lamp frame obscured by a multi-pinhole mask. The arc lines detected in the frame (top right image panel) are used to fit a global dispersion solution (middle right image panel). The residuals of the fits as compared to measured order-edge and arc-line locations can be found in the bottom panels.}
        \label{fig:soxspipe-results}
\end{figure}

The \ac{SOXS} \ac{E2E} simulator \cite{Genoni20} is capable of producing simulated 2D images in the \ac{SOXS} format that take into account the main optical behaviour of the system (grating dispersion, sampling, PSF, noises and position of various resolution elements coming from full ray-tracing). By using test-driven development throughout the development process, combined with ‘extreme’ mock data generated from the \ac{E2E} simulator, we can verify the pipeline is not only able to reduce a typical data set but also data that is far from ideal. This extreme data helps us push the pipeline to the limits of its capabilities and allows us to defensively develop against the edge-case scenarios the pipeline will most certainly experience at some point in production mode. Figure \ref{fig:soxspipe-results} gives an indication of the quality of the reductions achieved by \texttt{soxspipe} when reducing \ac{E2E} calibration frames.

\begin{table}[t]
\caption{Final data-products generated by the \ac{SOXS} pipeline.} 
\label{tab:products}
\begin{center}       
\begin{tabular}{|l|l|} 
\hline
\rule[-1ex]{0pt}{3.5ex}  Product  & Description         \\
\hline
\rule[-1ex]{0pt}{3.5ex}  1D Source Spectra  & \makecell[l]{1D spectra in FITS binary table format, one for each arm.  \\ Each FITS spectrum file will contain 4 extensions:  \\ 1. Wavelength- and flux-calibrated spectra with absolute flux correction \\ via scaling to acquisition image source photometry, \\ 2. an additional spectrum with correction for telluric absorption via \\ MOLECFIT, \\ 3. the variance array and \\ 4. the sky-background spectra.}      \\
\hline
\rule[-1ex]{0pt}{3.5ex} 1D Merged Source Spectrum & \makecell[l]{1D UV-VIS \& NIR merged spectrum in FITS binary table \\ format with PDF visualisation. This spectrum \\ will be rebinned to a common pixel scale for each arm. \\ This spectrum file will also have the same 4 \\ extensions described above.}  \\
\hline
\rule[-1ex]{0pt}{3.5ex} 2D Source Spectra  & \makecell[l]{A 2D FITS image for each spectral arm containing wavelength \\ and flux calibrated spectra (no other corrections applied) allowing \\ users to perform source extraction with their tool of choice. Note that \\ rectification of the curved orders in the NIR introduces a source \\ of correlated noise not present in extractions performed on the \\ unstraightened orders as done by the pipeline.}   \\
\hline
\rule[-1ex]{0pt}{3.5ex}  Acquisition Camera Images & \makecell[l]{\emph{ugrizy} astrometrically and photometrically (\emph{griz} only) calibrated to \\ Refcat2 \cite{2018ApJ...867..105T} } \\
\hline
\end{tabular}
\end{center}
\end{table} 

\begin{figure}[t]\centering
  \includegraphics[width=10cm]{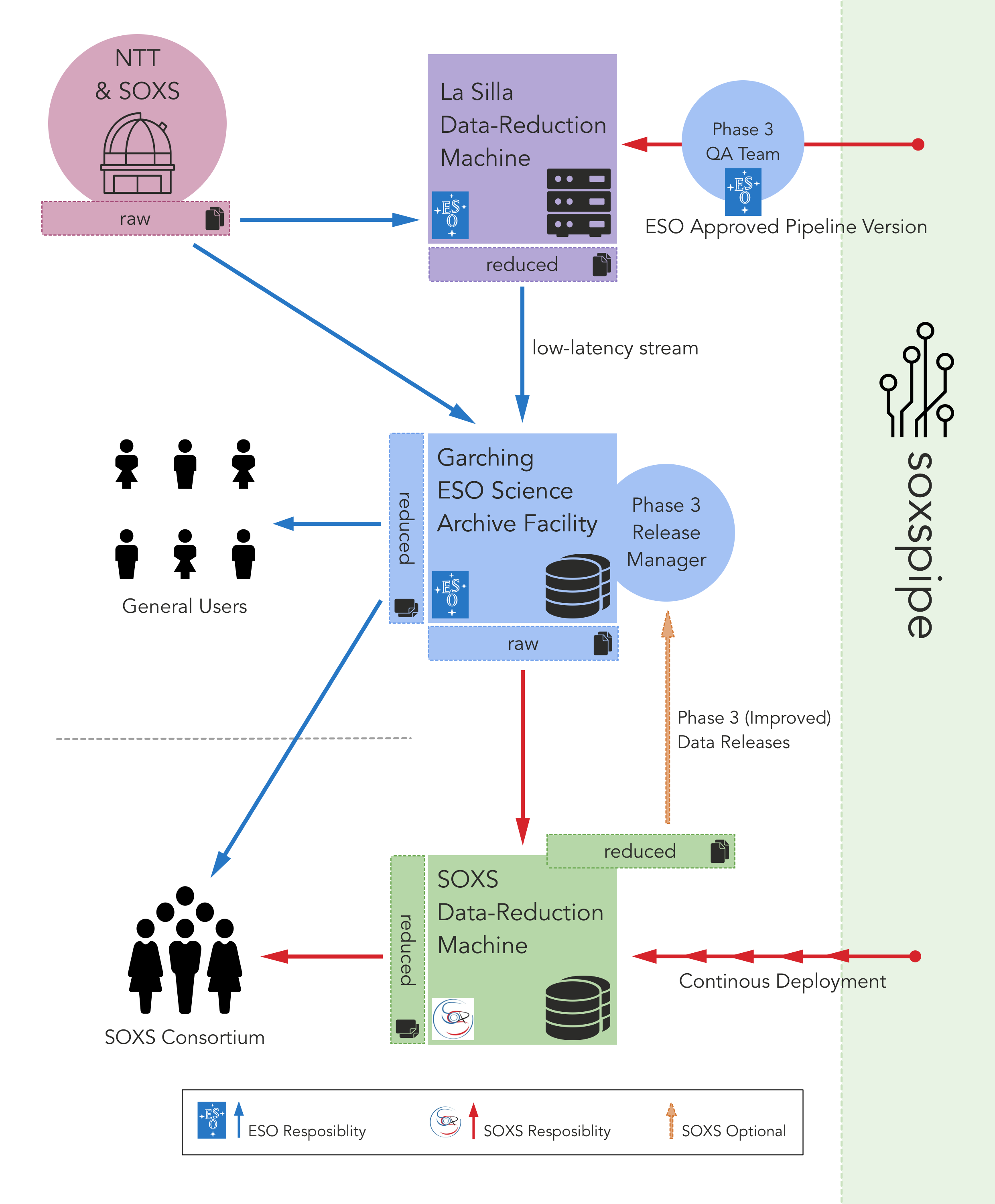}
  \caption{The proposed \ac{SOXS} data flow. Data is acquired by the \ac{NTT} and \ac{SOXS} on the summit of La Silla, Chile (top right). Raw data is reduced on the summit (top centre) and transferred within minutes to the \ac{ESO} \ac{SAF} (Garching, Germany) where data-right owners can access it (central in blue). In parallel, the \ac{SOXS} consortium will also reduce their data on a remote machine (probably cloud-based) with a leading-edge version of the pipeline (bottom in green). If at any point it is decided that new development of the pipeline has led to significantly improved data products compared to those hosted on the \ac{SAF} the consortium may opt for a complete reprocessing and replacement of the data on the \ac{SAF} via a dedicated Phase 3 Data Release (orange arrow).}
  \label{fig:dataflow}
\end{figure}

The pipeline code is open-source, hosted on Github\footnote{https://github.com/thespacedoctor/soxspipe} and connected to a Jenkins Continuous Integration/Continuous Deployment server via Github's webhooks. Any new push of code to a branch on the GitHub repository triggers a new 'build' of the code on the CI server where all unit tests are run. If all tests pass the branch can be merged into main development branch. If it is the main/production branch being tested, and all tests pass, then a new dot release version of the code is automatically shipped to PyPI\footnote{https://pypi.org/project/soxspipe/} and conda-forge ready for deployment.

\section{Data Products and Data Flow}
\label{sec:dataflow} 

\texttt{soxspipe} will reduce data into a set of final data products (see Table \ref{tab:products} for details) which shall meet \ac{ESO} Phase 3 standards `out-of-the-box'. This has the benefit of allowing us to build an automated workflow (see Figure \ref{fig:dataflow}) to reduce data directly on the La Silla summit immediately after the data is acquired by the \ac{NTT} and \ac{SOXS} and then stream the reduced data directly into the \ac{ESO} \ac{SAF} \cite{2011AAS...21830503R} in Garching, Germany. Owners of the data will then be able to access the fully-reduced data alongside the raw data within minutes of the shutter closing on their observation. This low-latency, automatic reduction is possible thanks to the fixed format of \ac{SOXS} (apart from exchangeable slit) allowing calibration frames to be prepared ahead of time before science data reductions. The \ac{SAF} then acts as both a data distribution solution and also fulfils the \ac{SOXS} consortium's legacy archive requirements.

Access to the `open stream' method of shipping reduced data directly to the \ac{ESO} \ac{SAF} will initially require the \ac{ESO} Archive Science Group to review and verify a moderately sized collection of \texttt{soxspipe} reduced data. Once the quality and content of the data produced by the pipeline have met \ac{ESO} Phase 3 standards we will then be allowed to ship data products to the archive without further need of passing through a gatekeeper. The pipeline will automatically reduce on all point-source targets above an AB magnitude of r = 19 (with the stretch goal of r = 20). For sources below this magnitude, the pipeline will attempt to automatically reduce the data but may require some user interaction to optimise object extraction.

\section{Conclusions}
\label{sec:conclusions} 

The \ac{SOXS} Pipeline \texttt{soxspipe} has been designed and written in object-orientated Python 3 using an agile framework of development. Built with core aims of allowing for fast, automatic reduction of raw data, streaming reduced data into the \ac{ESO} \ac{SAF} without need for human intervention, while also providing end-users with a simple-to-use, well-documented command-line interface, it is our hope that the pipeline will help facilitate the success of \ac{SOXS} in the years to come.

\bibliography{report} 

\begin{thebibliography}{10}

\bibitem{Schipani20}
Schipani, P. et~al., ``Development status of the {SOXS} spectrograph for the
  {ESO-NTT} telescope,'' {\em Proc. SPIE} {\bf 11447},  1144709 (2020).

\bibitem{Aliverti18}
Aliverti, M. et~al., ``The mechanical design of {SOXS} for the {NTT},'' {\em
  Proc. SPIE} {\bf 10702},  1070231 (2018).

\bibitem{Aliverti20}
Aliverti, M. et~al., ``Manufacturing, integration and mechanical verification
  of {SOXS},'' {\em Proc. SPIE} {\bf 11447},  114476O (2020).

\bibitem{Biondi18}
Biondi, F. et~al., ``The assembly integration and test activities for the new
  {SOXS} instrument at {NTT},'' {\em Proc. SPIE} {\bf 10702},  107023D (2018).

\bibitem{Biondi20}
Biondi, F. et~al., ``The {AIV} strategy of the common path of {S}on of
  {X-S}hooter,'' {\em Proc. SPIE} {\bf 11447},  114476P (2020).

\bibitem{Brucalassi18}
Brucalassi, A. et~al., ``The acquisition camera system for {SOXS} at {NTT},''
  {\em Proc. SPIE} {\bf 10702},  107022M (2018).

\bibitem{Brucalassi20}
Brucalassi, A. et~al., ``Final design and development status of the acquisition
  and guiding system for {SOXS},'' {\em Proc. SPIE} {\bf 11447},  114475V
  (2020).

\bibitem{Capasso18}
Capasso, G. et~al., ``{SOXS} control electronics design,'' {\em Proc. SPIE}
  {\bf 10707},  107072H (2018).

\bibitem{Claudi18}
Claudi, R. et~al., ``The common path of {SOXS} ({S}on of {X}-{S}hooter),'' {\em
  Proc. SPIE} {\bf 10702},  107023T (2018).

\bibitem{Claudi20}
Claudi, R. et~al., ``Operational modes and efficiency of {SOXS},'' {\em Proc.
  SPIE} {\bf 11447},  114477C (2020).

\bibitem{Colapietro20}
Colapietro, M. et~al., ``Progress and tests on the {I}nstrument {C}ontrol
  {E}lectronics for {SOXS},'' {\em Proc. SPIE} {\bf 11452},  1145225 (2020).

\bibitem{Cosentino18}
Cosentino, R. et~al., ``The vis detector system of {SOXS},'' {\em Proc. SPIE}
  {\bf 10702},  107022J (2018).

\bibitem{Cosentino20}
Cosentino, R. et~al., ``Development status of the {UV-VIS} detector system of
  {SOXS} for the {ESO-NTT} telescope,'' {\em Proc. SPIE} {\bf 11447},  114476C
  (2020).

\bibitem{Genoni20}
Genoni, M. et~al., ``{SOXS} {E}nd-to-{E}nd simulator: development and
  applications for pipeline design,'' {\em Proc. SPIE} {\bf 11450},  114501B
  (2020).

\bibitem{Kuncarayakti20}
Kuncarayakti, H. et~al., ``Design and development of the {SOXS} calibration
  unit,'' {\em Proc. SPIE} {\bf 11447},  1144766 (2020).

\bibitem{Ricci18}
Ricci, D. et~al., ``Architecture of the {SOXS} instrument control software,''
  {\em Proc. SPIE} {\bf 10707},  107071G (2018).

\bibitem{Ricci20}
Ricci, D. et~al., ``Development status of the {SOXS} instrument control
  software,'' {\em Proc. SPIE} {\bf 11452},  114522Q (2020).

\bibitem{Rubin18}
Rubin, A. et~al., ``{MITS}: the {M}ulti-{I}maging {T}ransient {S}pectrograph
  for {SOXS},'' {\em Proc. SPIE} {\bf 10702},  107022Z (2018).

\bibitem{Rubin20}
Rubin, A. et~al., ``Progress on the {UV-VIS} arm of {SOXS},'' {\em Proc. SPIE}
  {\bf 11447},  114475L (2020).

\bibitem{Sanchez18}
Sanchez, R.~Z. et~al., ``Optical design of the {SOXS} spectrograph for {ESO
  NTT},'' {\em Proc. SPIE} {\bf 10702},  1070227 (2018).

\bibitem{Sanchez20}
Sanchez, R.~Z. et~al., ``{SOXS}: effects on optical performances due to gravity
  flexures, temperature variations, and subsystems alignment,'' {\em Proc.
  SPIE} {\bf 11447},  114475F (2020).

\bibitem{Schipani16}
Schipani, P. et~al., ``The new {SOXS} instrument for the {ESO NTT},'' {\em
  Proc. SPIE} {\bf 9908},  990841 (2016).

\bibitem{Schipani18}
Schipani, P. et~al., ``{SOXS}: a wide band spectrograph to follow up
  transients,'' {\em Proc. SPIE} {\bf 10702},  107020F (2018).

\bibitem{Vitali18}
Vitali, F. et~al., ``The {NIR} spectrograph for the new {SOXS} instrument at
  the {NTT},'' {\em Proc. SPIE} {\bf 10702},  1070228 (2018).

\bibitem{Vitali20}
Vitali, F. et~al., ``The development status of the {NIR} arm of the new {SOXS}
  instrument at the {ESO/NTT} telescope,'' {\em Proc. SPIE} {\bf 11447},
  114475N (2020).

\bibitem{Young20}
Young, D. et~al., ``The {SOXS} data reduction pipeline,'' {\em Proc. SPIE} {\bf
  11452},  114522D (2020).

\bibitem{Marty22}
Marty, L. et~al., ``The quality check system architecture for soxs,'' {\em
  Proc. SPIE} (in press 2022).

\bibitem{Landoni22}
Landoni, M. et~al., ``The soxs scheduling system,'' {\em Proc. SPIE} (in press
  2022).

\bibitem{2018ApJ...867..105T}
{Tonry}, J.~L., {Denneau}, L., {Flewelling}, H., {Heinze}, A.~N., {Onken},
  C.~A., {Smartt}, S.~J., {Stalder}, B., {Weiland}, H.~J., and {Wolf}, C.,
  ``{The ATLAS All-Sky Stellar Reference Catalog},'' {\em \apj}~{\bf 867},  105
  (Nov. 2018).

\bibitem{2011AAS...21830503R}
{Romaniello}, M., ``{ESO's Science Archive Facility},'' in [{\em American
  Astronomical Society Meeting Abstracts \#218}{\nolinebreak\hspace{0.1em}]},
  {\em American Astronomical Society Meeting Abstracts} {\bf 218},  305.03 (May
  2011).

\end{thebibliography}
\bibliographystyle{spiebib} 

\end{document}